\documentclass[%
 reprint,
 superscriptaddress,
 amsmath,amssymb,
 aps,
 prl,
 floatfix
]{revtex4-2}

\usepackage{graphicx}
\usepackage{dcolumn}
\usepackage{bm}
\usepackage{hyperref}

\usepackage[usenames,dvipsnames]{color}
\usepackage{colortbl}
\newcommand{\REV}[1]{#1} 
\newcommand{\REVFINAL}[1]{#1}

\newcommand{\DEL}[1]{}

\begin{document}


\title{Self-trapping of slow electrons in the energy domain}

\author{Maor Eldar}
\affiliation{Department of Physics, Technion---Israel Institute of Technology, Haifa 32000, Israel}
\affiliation{Solid State Institute, Technion---Israel Institute of Technology, Haifa 32000, Israel}
\affiliation{\REV{The Helen Diller Quantum Center, Technion---Israel Institute of Technology, Haifa 32000, Israel}}

\author{\REV{Zhaopin Chen}}
\affiliation{Department of Physics, Technion---Israel Institute of Technology, Haifa 32000, Israel}
\affiliation{Solid State Institute, Technion---Israel Institute of Technology, Haifa 32000, Israel}
\affiliation{\REV{The Helen Diller Quantum Center, Technion---Israel Institute of Technology, Haifa 32000, Israel}}

\author{Yiming Pan}
\affiliation{Department of Physics, Technion---Israel Institute of Technology, Haifa 32000, Israel}
\affiliation{\REV{The Helen Diller Quantum Center, Technion---Israel Institute of Technology, Haifa 32000, Israel}}
\affiliation{School of Physical Science and Technology and Center for Transformative Science, ShanghaiTech University, Shanghai 200031, China}

\author{Michael Kr\"uger}
\altaffiliation{Corresponding author: krueger@technion.ac.il}
\affiliation{Department of Physics, Technion---Israel Institute of Technology, Haifa 32000, Israel}
\affiliation{Solid State Institute, Technion---Israel Institute of Technology, Haifa 32000, Israel}
\affiliation{\REV{The Helen Diller Quantum Center, Technion---Israel Institute of Technology, Haifa 32000, Israel}}

\date{\today}

\begin{abstract}
{\bf The interaction of light and swift electrons has enabled phase-coherent manipulation and acceleration of electron wavepackets. Here we investigate this interaction in a new regime where low-energy electrons ($\sim$20-200\,eV) interact with a phase-matched light field. Our analytical and \REVFINAL{one-dimensional} numerical study shows that slow electrons are subject to strong confinement in the energy domain due to the non-vanishing curvature of the electron dispersion. The spectral trap is tunable and an appropriate choice of light field parameters can reduce the interaction dynamics to only two energy states. The capacity to trap electrons expands the scope of electron beam physics, free-electron quantum optics and quantum simulators.}
\end{abstract}

\maketitle

The interaction between free electrons and light, resulting in a high degree of coherent control of the electron wavefunction, has been intensively studied during the past two decades~\cite{Barwick2009,GarciaDeAbajo2010,Kirchner2014,Feist2015,Dahan2020,Henke2021}. \REV{While energy-momentum conservation in this interaction cannot be fulfilled in free space, introducing a third medium or body allows for the exchange of energy between photons and electrons. This} can be achieved using different approaches, such as photon-induced near-field electron microscopy (PINEM,~\cite{Barwick2009,Feist2015,Shiloh2022}), optical field discontinuities at interfaces~\cite{Kirchner2014,Morimoto2018}, dielectric laser acceleration (DLA,~\cite{Breuer2013,Peralta2013,Adiv2021}), flat surfaces with phase-matched near-fields~\cite{Kozak2017a,Dahan2020}, photonic cavities~\cite{Kfir2020,Henke2021} or ponderomotive acceleration~\cite{Kozak2018,Huang2021,Tsarev2023}. The main signature of the interaction is the appearance of sidebands in the electron energy spectrum \REV{that are spaced by photon energy quanta of the driving light as a result of energy and momentum transfer between electrons and light (see, e.g.,~\cite{Barwick2009,Feist2015,Adiv2021,Dahan2021,Tsarev2023}). The sidebands enable} attosecond electron pulses~\cite{Priebe2017,Morimoto2018,Kozak2019,Black2019,Schonenberger2019,Morimoto2020,Sears2008a}. All these works have employed fast electrons (10-200\,keV), enabling a straightforward understanding of much of the physics, primarily in the simplified picture of multi-level Rabi oscillations~\cite{Feist2015}. Here, the interaction with the optical field allows the electron to perform a quantum random walk on an infinite energy ladder. The multi-level Rabi oscillations model is based on approximations for fast electrons which include neglecting the electron momentum recoil (non-recoil approximation), applying the short-time interaction approximation without phase-matching, and neglecting the ponderomotive forces exerted by the strong light field. However, recent theory works have begun focusing on strong-field slow-electron interactions, for example via numerical study of strong off-resonant coupling at $\sim$0.1-1\,keV~\cite{Talebi2020}\REV{,} inelastic ponderomotive scattering at $\sim$10\,keV~\cite{kozak2022asynchronous} \REV{and Jaynes-Cummings-type interactions with a cavity~\cite{Karnieli2023}}. Recent advancements in low-energy electron microscopy and source development \REV{have enabled the generation of tailored low-energy electron pulses~\cite{Hommelhoff2006,Ropers2007,Barwick2007,Kruger2011,Quinonez2013,Wimmer2014,Ehberger2015,Vogelsang2018,Vogelsang2018a,Kruger2018,Eldar2022}, opening up first experimental studies of electron-light interactions in this novel regime}.

In this Letter, we perform a theory study of the phase-matched interaction of slow electrons ($\sim$20-200\,eV) with a strong optical field. We find that resonant interactions in this regime cause a strong confinement of the low-energy electron spectrum due to the non-vanishing curvature of the electron dispersion. The latter acts as a quadratic trapping potential in the energy domain, setting a limit to the quantum random walk of the electrons. We show that this trapping in the energy domain is tunable due to the competition of energy ladder hopping and electron dispersion which depends strongly on the frequency and strength of the optical driving field. Our findings demonstrate that the rich toolbox for manipulating free electrons with light is not restricted to fast electrons, but gives rise to interesting new phenomena in the regime of low-energy electrons, much beyond the multi-level Rabi picture.

\begin{figure}[htb!]
\begin{centering}
\includegraphics[width=\columnwidth]{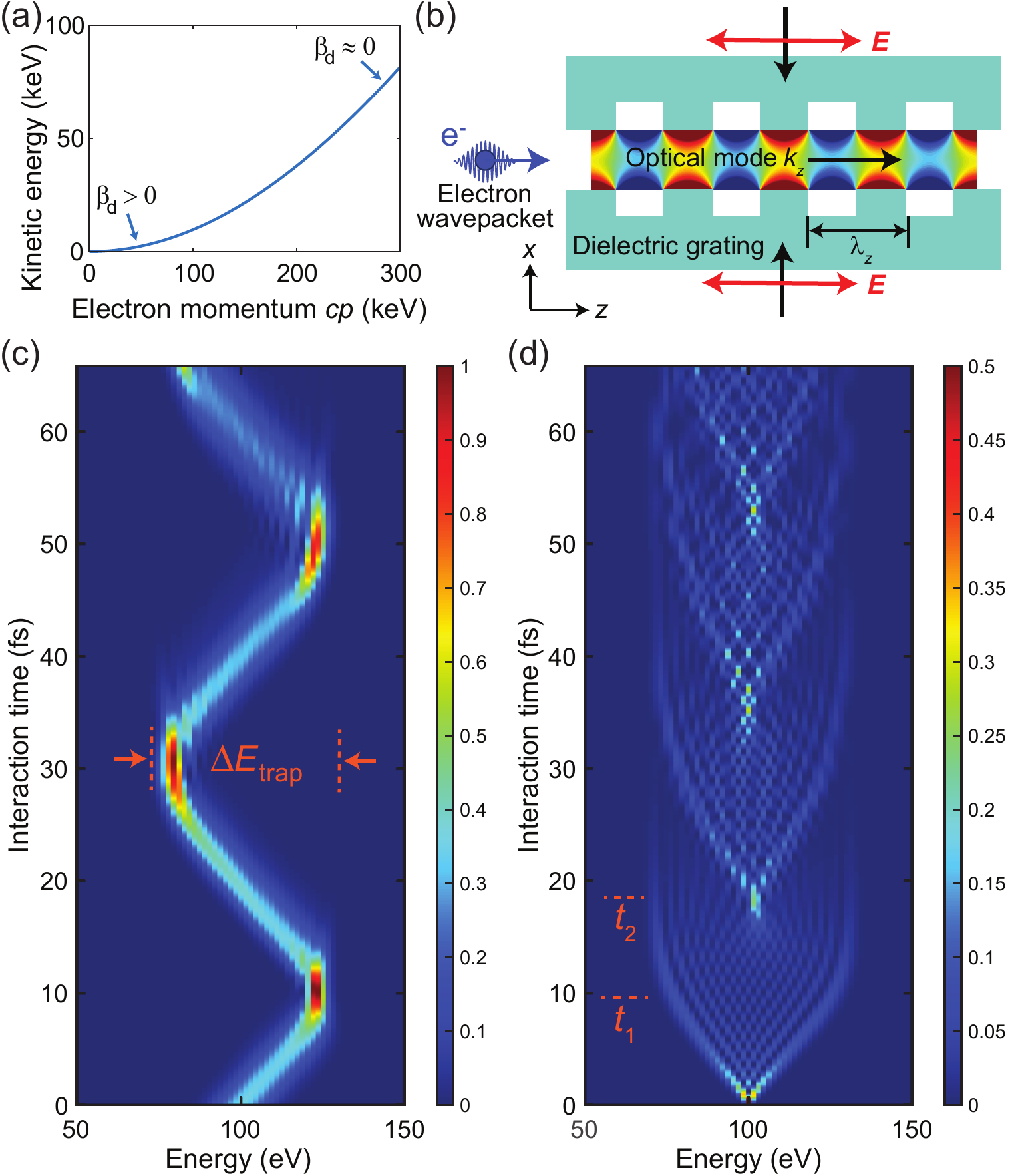}
\par\end{centering}
\caption{Phase-matched low-energy electron light-matter interactions. (a) The electron dispersion for low-energy electrons displays a strong curvature \REV{$\beta_\mathrm{d}$}. (b) Sketch of a possible \REV{experimental }setup. An electron is entering into a region of interaction with a phase-matched optical \REV{mode with wavevector $k_z$}. Phase-matching is achieved via \REV{a mirror-symmetric pair of }dielectric gratings\REV{ with a period of $\lambda_z$ illuminated from both sides}. (c) Evolution of the electron spectrum as a function of interaction time\REV{, as obtained from Eq.~\ref{eq1}}. The initial state is a localized wavepacket with an energy width corresponding to five photons $\Delta E\sim 7.5$\,eV \REV{(438\,as temporal duration)}. \REV{The dashed lines represent the trap edges and the trap width $\Delta E_\mathrm{trap}$.} (d) The same for a plane wave. In (c) and (d), the initial kinetic energy is set to $E_{0}=100$\,eV. The optical field amplitude is \REV{$E_\mathrm{f}=0.5\,\mathrm{V\,nm}^{-1}$}, the photon energy is $\hbar \omega=1.54\,$eV and the phase is $\phi_{0}=0$.}
\label{fig1}
\end{figure}

In order to understand the strong-field dynamics of slow electrons, we perform a Floquet-Bloch analysis of the \DEL{one-dimensional }time-dependent Schr{\"o}dinger equation\REV{ (TDSE)}. \REV{We reduce the problem to one spatial dimension along the electron's propagation direction (see Supplemental Material~\cite{NoteSuppl}\nocite{Li2023,Soong2012,Korpel1981} for a full derivation and justification of this reduction).} Key to our analysis is the electron's Hamiltonian given by $H_{0}=E_{0}+v_{0}(p-p_{0})+{(p-p_{0})^{2}}/{(2\gamma^{3}m)}$, which we retrieve from expanding the relativistic dispersion to second order around the initial electron momentum $p_0$. \REV{The full relativistic dispersion is illustrated in Fig.~\ref{fig1}(a). It displays a non-vanishing curvature $\beta_\mathrm{d}$ for low energies, which necessitates the second-order expansion for $H_0$.} Here $m$ is the electron rest mass and $E_{0}=\sqrt{m^2c^4+p_{0}^2c^2} - mc^2$, $p_{0}=\gamma m v_{0}$ and $v_{0}$ are the initial kinetic energy, momentum and velocity, respectively\REV{, of the electron, while} $\gamma$ is the Lorentz factor and $c$ the speed of light. \REV{In our proposed setup, we assume an optical mode with wavevector $k_z$, where $k_z$ can be chosen such that a phase-matched electron-light interaction is achieved. An example of this is an evanescent mode induced at a DLA-type double grating structure with period $\lambda_z$~\cite{Peralta2013}, where $k_z = 2\pi/\lambda_z$ (see Fig.~\ref{fig1}(b) and Supplemental Materials~\cite{NoteSuppl} for a detailed discussion). We neglect other modes in our theory study since they cannot lead to a net modulation of the electron.} The interaction Hamiltonian is given by $H_\mathrm{I}=-\frac{e}{2\gamma m}(A\cdot p+p\cdot A)+\frac{e^2A^2}{2\gamma m}$, where $A$ is the optical field's vector potential and $e$ the elementary charge. For a monochromatic field, the vector potential is given by $A(z,t)=-\frac{E_\mathrm{f}}{\omega}\sin(\omega t-k_{z}z+\phi_{0})$, where $E_\mathrm{f}$ \REV{and $\omega$ are the electric field amplitude and angular frequency, respectively. $\phi_{0}$ is the initial phase of the electric field at the beginning of the interaction ($t = 0$).}

The electron-light interaction is imprinted on the resulting energy spectrum. After the electron enters the interaction region where the optical field is present it absorbs or emits an integer number of photons from the field, resulting in equally spaced energy states. Hence, encapsulating this spatiotemporal periodicity we employ the Floquet-Bloch ansatz $|\psi(t)\rangle={\sum_{n=-\infty}^{\infty}}a_{n}(t)e^{-i\omega_{n}t}|k_{n}\rangle$ where $n$ denotes the number of photons absorbed ($n>0$) or emitted ($n<0$) and $a_{n}$ are the probability amplitudes corresponding to the respective energy-momentum states $|k_{n}\rangle$. Here $k_{n}=k_{0}+nk_{z}$ and $\omega_{n}=\omega_{0}+n\omega$ denote the wavenumber and angular frequency values, respectively. The electron's initial kinetic energy and momentum are given by $\hbar \omega_{0} \equiv (\gamma - 1) m c^2$ and $\hbar k_{0} \equiv \gamma mv_{0}$, respectively. Substituting the ansatz into the Schr{\"o}dinger equation we find

\begin{equation}
\begin{split}\label{eq1}
i\hbar\frac{\partial a_{n}(t)}{\partial t}=-\left[n\hbar(\omega-v_{0}k_{z})-n^{2}\frac{(\hbar k_{z})^{2}}{2m}+U_\mathrm{p}\right]a_{n}(t) \\
+\delta\left((k_{n+1}-\frac{k_{z}}{2})a_{n+1}(t)e^{i\phi}+(k_{n-1}+\frac{k_{z}}{2})a_{n-1}(t)e^{-i\phi}\right) \\
-\frac{U_\mathrm{p}}{2}\left(a_{n+2}(t)e^{i2\phi_{0}}+a_{n-2}(t)e^{-i2\phi_{0}}\right),
\end{split}
\end{equation}

where $U_\mathrm{p}=e^{2}E_\mathrm{f}^{2}/{(4m\omega^{2})}$ is the ponderomotive energy, $\delta=e\hbar E_\mathrm{f}/(2m\omega)$ and $\phi=\phi_{0}+\frac{\pi}{2}$. Eq.~\ref{eq1} reveals several intriguing and unique characteristics of the slow electron-light interaction, as we will explain below. 

In order to resonantly couple the slow electron and the field, we require the conservation of both energy and momentum for single photon emission or absorption by the electron at the quantum level, resulting in the phase-matching condition of equal electron (group) velocity and optical field (phase) velocity~\cite{park2010photon}. This phase-matching condition also holds true for low-energy electrons (see Supplemental Material~\cite{NoteSuppl}). \REV{In a DLA realization of our scheme using a grating, phase matching requires setting the grating period in accordance with the electron velocity ($v=\beta c$) and the impinging field's wavelength $\lambda_\mathrm{f}$, namely $\lambda_z=\beta \lambda_\mathrm{f}$~\cite{Breuer2013}.} We note that it is challenging to achieve phase-matching for slow electrons ($E_0 \approx 100\,$eV) since it requires a significant reduction of the phase velocity of the optical field, but it is not impossible. For instance, DLA requires a grating structure with a period \REV{$\lambda_z \sim 16$\,nm} for 100\,eV electrons and 800\,nm \REV{free-space light} wavelength\REV{, which is in reach of today’s technological capabilities. As an alternative to DLA-type gratings,} the use of plasmonic metamaterials~\cite{Tsakmakidis2017} or materials with very high refractive index $(n\approx50)$, such as $\mathrm{SrTiO}_{3}$ at low temperatures \REV{($\sim$10\,K, experimentally realized in~\cite{sakudo1971dielectric})}, can generate \REV{evanescent} waves with an extremely small phase velocity. \REV{A prism configuration incorporating such materials enables phase-matched interactions~\cite{Dahan2021}. Laser-triggered needle tips~\cite{Hommelhoff2006,Ropers2007,Barwick2007,Kruger2011,Quinonez2013,Wimmer2014,Ehberger2015,Vogelsang2018,Vogelsang2018a,Kruger2018} produce suitable electron pulses with tunable bandwidth~\cite{Wimmer2014}.}

The $A^2$ ponderomotive term of the interaction Hamiltonian is reflected in the third line of Eq.~\ref{eq1}, permitting two-photon exchanges in each interaction event. It becomes more important for slow electrons since the competing linear interaction term $p\cdot A$ decreases with velocity. However, even for a 50\,eV electron, strong electric fields with amplitudes on the order of $10\,\mathrm{V\,nm}^{-1}$ have to be applied to make the ponderomotive term comparable to the $p\cdot A$ term. In contrast to free-space ponderomotive schemes~\cite{freimund2001observation,Kozak2018,kozak2022asynchronous}, material damage thresholds limit the applicable field intensity. Under this limitation, a strong ponderomotive effect can still be reached if the electron kinetic energy approaches the single photon energy, $eE_\mathrm{f}/\omega \approx p_0$ (see Supplemental Material~\cite{NoteSuppl}).

The second characteristic of slow-electron strong-field interaction is the breakdown of the non-recoil approximation. For fast electrons, the detuning in momentum recoil owing to photon absorption and emission may be neglected, as explicitly represented by $k_{n+1}\approx k_{n-1}\approx k_{0}$~\cite{Feist2015}. However, this is no longer true for slow electrons. In order to reveal this, we focus on the ratio ${|\Delta v|}/{v_{0}}$, where $\Delta v$ is the velocity change caused by a single photon exchange. \REV{This quantity does not depend on the field strength.} For an electron with $E_{0}=50$\,eV and a photon energy $\hbar\omega=1.54$\,eV, we find ${|\Delta v|}/{v_{0}}=\sqrt{{\hbar\omega}/{E_{0}}}\approx10^{-1}$. In contrast, for a fast electron of $E_{0}=200$\,keV we find ${|\Delta v|}/{v_{0}} \approx 10^{-3}$, hence the detuning can be neglected. The breakdown of the non-recoil approximation amounts to an effective symmetry breaking between absorption and emission, which is most pronounced at higher orders of photon scattering and manifests as a slightly altered coupling constant at the level of single photon exchange. In addition, we find an asymmetric evolution pattern in the electron spectrum when electrons and light are not phase-matched. This leads to a different group velocity dispersion for emission and absorption sidebands, resulting in asymmetric Bloch-type oscillations (see Supplemental Material~\cite{NoteSuppl}).

The first line of Eq.~\ref{eq1} reveals two potentials acting on the time evolution of the sideband momentum states, $n\hbar(\omega-v_{0}k_{z})$ and $-{(n\hbar k_{z})^{2}}/({2m})$. The first potential term is linear in $n$ and results from the mismatch between electron group velocity and light field phase velocity. Under the condition $\omega \neq v_{0}k_{z}$, this phase mismatch acts effectively as a linear potential on a synthetic frequency space resulting in Bloch oscillation dynamics in the energy spectrum as reported previously for fast free electrons~\cite{pick2018bloch,pan2022synthetic}, which lead to Wannier-Stark localization~\cite{wannier1960wave}. However, when we introduce phase-matching through $\omega = v_{0}k_{z}$, the linear term vanishes and we are left with the quadratic potential term. The latter is only significant for slow electrons 
and acts as a confining potential for the electron's spectral evolution. Figures~\ref{fig1}(c) and (d) displays the population probability of the sideband momentum states as function of interaction time obtained from numerical solutions of~Eq.~\ref{fig1}. We show the results for two initial conditions, a localized Gaussian wavepacket (Fig.~\ref{fig1}(c)) and a single energy (plane wave, Fig.~\ref{fig1}(d)), both centered at $100$\,eV initial kinetic energy. The field strength is set to \REV{$E_\mathrm{f}=0.5\,\mathrm{V\,nm}^{-1}$} and the single photon energy is $\hbar\omega=1.54$\,eV. We observe a spectral evolution exhibiting strong confinement and oscillations. For the electron wavepacket, we find that the spectral evolution follows a classical Lorentz-like trajectory for a charged particle in an electric field. In contrast, for a plane-wave electron we see spectral broadening to a superposition state, which subsequently ceases to expand and suddenly collapses to a single sideband. In both cases, we observe a spectral asymmetry in the energy spectrum as a small but noticeable difference ($\Delta n\approx3$) between the maximally populated energy states of absorption and emission. This asymmetry is a signature of the non-vanishing curvature of the electron dispersion at low energies and accumulates across successive photon scattering events. \REV{The validity of the results of our Floquet-Bloch model given by Eq.~\ref{eq1} is confirmed by a numerical solution of the full TDSE (see Supplemental Material~\cite{NoteSuppl}).}

\begin{figure}[htb!]
\begin{centering}
\includegraphics[width=\columnwidth]{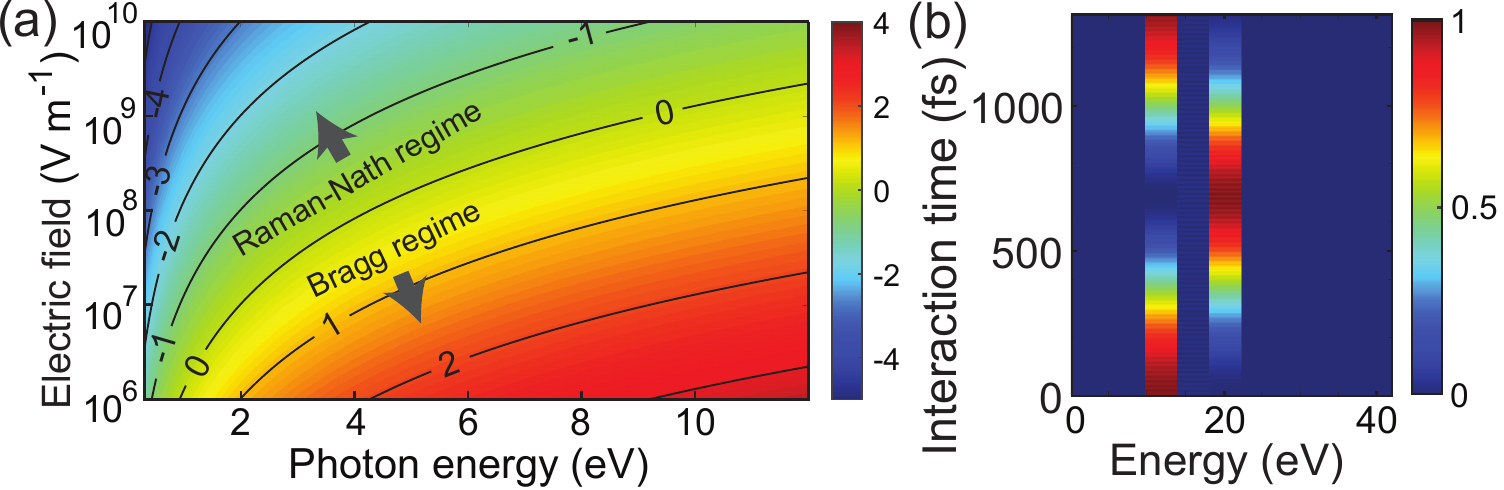}
\par\end{centering}
\caption{(a) Transition between Raman-Nath and Bragg regimes. We show $\log \rho$ as a function of photon energy $\hbar\omega$ and electric field strength $E_\mathrm{f}$. The initial electron energy is 100\,eV. (b) Spectral evolution in the Bragg regime \REV{following Eq.~\ref{eq1} for an initial \REVFINAL{off-resonant electron }energy of 12\,eV, photon energy $\hbar\omega = 4$\,eV and $E_\mathrm{f} = 1 \times 10^8\,\mathrm{V\,nm}^{-1}$ ($\rho \approx 10$, \REVFINAL{$\lambda_z \sim 3$\,nm for phase matching to 16\,eV electron energy, which corresponds to $n = 0$}). We observe Rabi-like oscillations between 12\,eV ($n = -1$) and 20\,eV ($n = +1$).}}
\label{fig2}
\end{figure}

In order to analytically capture the basic underlying physics in Fig.~\ref{fig1}, we lay aside the asymmetry in absorption and emission by applying the approximation $k_{n+1}\approx k_{n-1}\approx k_{0}$ and neglect ponderomotive effects. We are thus left with a simplified equation,
\begin{equation}
\label{eq2}
i\hbar\frac{\partial a_{n}}{\partial t}=\REV{\beta_\mathrm{d}} n^{2}a_{n}+\kappa a_{n+1}+\kappa^{*}a_{n-1},    
\end{equation}
where $\REV{\beta_\mathrm{d}}=\frac{(\hbar k_{z})^{2}}{2m}$ and $
\kappa=\frac{eE_\mathrm{f}\hbar k_{0}}{2m\omega}e^{i(\phi_{0}+\pi/2)}$. By relating \REV{$\beta_\mathrm{d}$} to the electron and light field parameters under phase-matching, we find $\REV{\beta_\mathrm{d}}\approx\frac{1}{4}{(\hbar\omega)^2}/{E_{0}}$. From Eq.~\ref{eq2} we can easily read the two competing processes that govern the spectral dynamics, namely harmonic oscillations and hopping. In analogy to solid-state terminology, we refer to the two terms on the r.h.s.\ of Eq.~\ref{eq2} as quadratic on-site potential ($n^2\REV{\beta_\mathrm{d}}$) and nearest-neighbor hopping ($\kappa$). Both parameters depend on the optical field parameters and the electron's energy dispersion. We note that the evolution of the electron spectrum governed by Eq.~\ref{eq2} is analogous to the Schr{\"o}dinger equation for atomic diffraction~\cite{meystre2021quantum}\REV{,} the transversal Kapitza-Dirac effect for electron diffraction~\cite{batelaan2007colloquium} \REV{and acousto-optics (see Supplemental Material~\cite{NoteSuppl})}. In order to distinguish the relative contribution of the two effects we invoke the Nath parameter, $\rho={\REV{\beta_\mathrm{d}}}/{\kappa}$~\cite{nath1938diffraction}. In Fig.~\ref{fig2}(a), we show $\log \rho$ as a function of photon energy and field strength. For $\rho \ll 1$, the hopping dominates, indicating the Raman-Nath regime, whereas for $\rho \gg 1$ we find the Bragg regime~\cite{moharam1978criterion}. In the following, we will discuss both regimes and their effects on electron spectral evolution. 

We first discuss the Raman–Nath regime ($\rho \ll 1$) which is reached when the hopping energy is much larger than the on-site energy. We can thus treat the latter as a perturbation. For instance, a slow electron with kinetic energy of 100\,eV can reach this regime while interacting with an electric field with wavelength $800$\,nm and  $E_\mathrm{f}=1\,\mathrm{V\,nm}^{-1}$, yielding $\rho \sim 10^{-3}$. The unperturbed equation then reads: $i\hbar\frac{\partial a_{n}(t)}{\partial t}=\kappa a_{n+1}(t)+\kappa^{*}a_{n-1}(t)$, which has a known analytical solution in the form of a Bessel function of the first kind, $a_{n}(t)=J_{n}(\frac{2|\kappa|}{i\hbar}t)e^{in\phi_{0}}$ (see Supplemental Material~\cite{NoteSuppl}). We obtain the probability $P_{n}(t)$ of locating the electron at a certain sideband as $P_{n}(t)=|J_{n}(\frac{2|\kappa|}{i\hbar}t)|^{2}$. The quantum coherent dynamics given by the Bessel function solution entail a symmetric diffraction pattern, as observed in \REV{Fig.~\ref{fig1}(d)} for times smaller than \REV{$t_1 \sim 9$\,fs}, in full analogy to PINEM with fast electrons~\cite{Feist2015}. However, at later times\REV{, up to $t_2 \sim 19$\,fs}, the rapid expansion is considerably slowed down as the confining term $(\propto n^2)$ becomes significant at these high sidebands. Physically, this is the result of energy–momentum conservation being violated at high photon scattering orders, due to the growing phase mismatch between light (linear dispersion) and free electron (quadratic dispersion). As a result, all \REV{electron trajectories are deflected from the trap edges and }self-collapse in close vicinity of the initial momentum state at around \REV{$t_2 \sim 19$\,fs, creating complex interference structures. This} evolution pattern then repeats periodically throughout the interaction time\REV{ due to the quadratic on-site potential}.

For $\rho \gg 1$ where the hopping is smaller \REV{than} the on-site potential term, we find the Bragg regime~\cite{batelaan2007colloquium}. As shown in Fig.~\ref{fig2}(a), entering the Bragg regime necessitates relatively weak electric fields ($\sim 10^{6}-10^{8}\,\mathrm{V\,m}^{-1}$) and high-energy photons ($\hbar \omega \sim 4-20$\,eV). In this regime, fewer sideband orders are populated compared to the Raman–Nath regime, which can be understood as the quantum electron optics analog of optical Bragg diffraction but occurring here in the electron spectrum. The on-site term $\propto n^2$ invokes a symmetry between photon emission and absorption scattering orders ($n,-n$). Therefore, even with a small hopping term of the Bragg regime, there exists a non-negligible coupling between these sidebands. For an initial spectrum containing only a single sideband, the Bragg regime can offer the strongest possible confinement involving only two sidebands, i.e., a coherent splitting of the electron spectrum much like a diffraction grating in optics. The dynamics takes the form of periodic oscillation between the two energy components, an effect known in neutron diffraction as Pendell{\"o}sung oscillations~\cite{Shull1968}, and was also observed in atom optics experiments~\cite{martin1988bragg}. A pure quantum effect occurs if the photon energy is on the order of the electron energy. Here, the slow electron populates only a few sidebands within a strong spectral confinement. The quantization grid in the electron phase space is comparable to its volume, therefore already exchanging a small number of photons leads to a large energy-momentum violation (large phase mismatch) and therefore eliminates higher sidebands (see Fig.~\ref{fig2}(b) for an example). Interestingly, for free-electron lasers a similar transition to a quantum regime was found~\cite{kling2015defines}.\REV{ We note that the truncation down to a two-level system with Rabi-like oscillations is only possible in the phase-matched Bragg regime and cannot be reached by other means, such as strong phase mismatch~\cite{pick2018bloch} or ponderomotive schemes~\cite{Huang2021,kozak2022asynchronous} (see Supplemental Material~\cite{NoteSuppl}).}

\begin{figure}[htb!]
\begin{centering}
\includegraphics[width=\columnwidth]{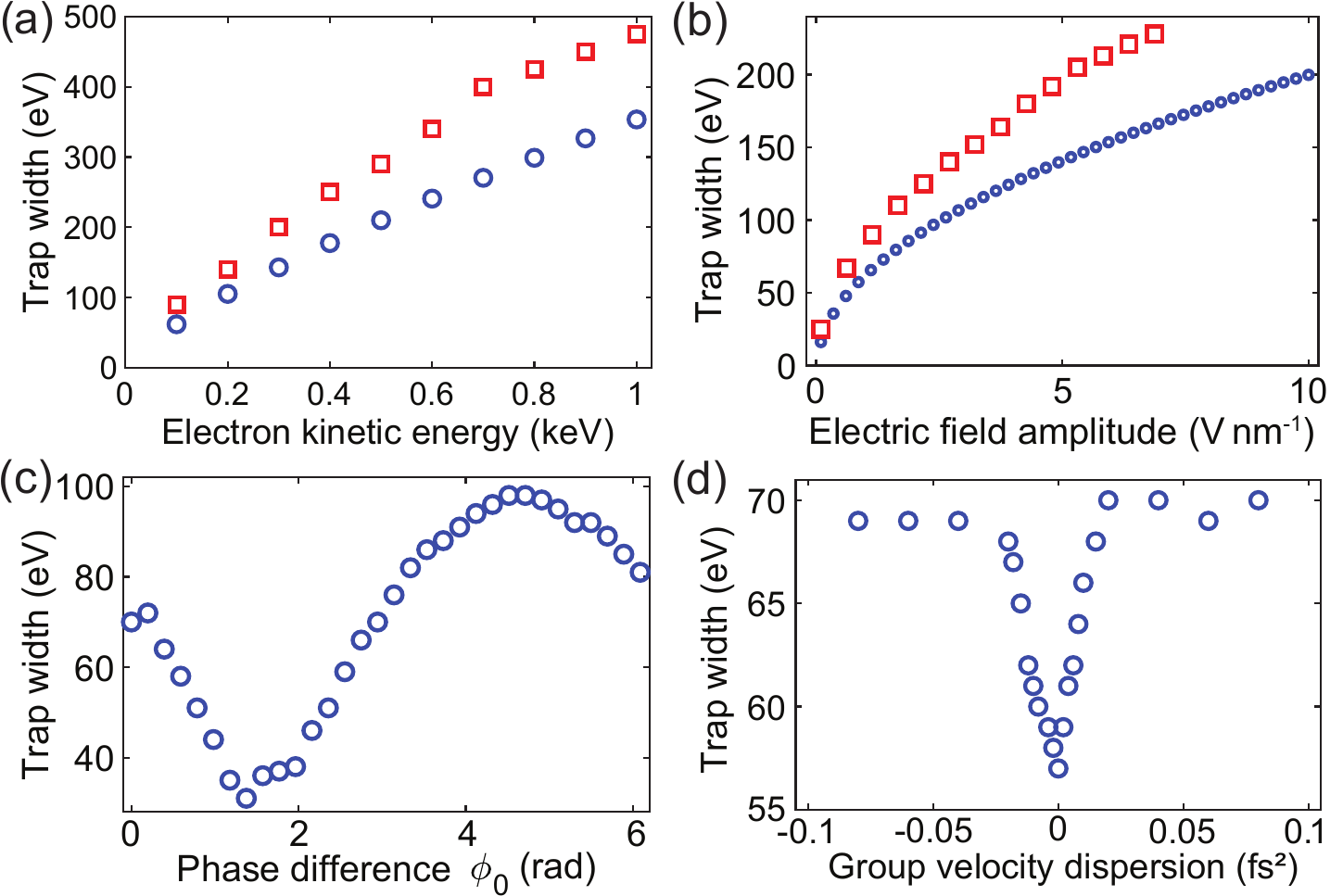}
\par\end{centering}
\caption {Parameter dependence of the trap width. (a) Dependence on electron energy for $E_\mathrm{f}=1\,\mathrm{V\,nm}^{-1}$, $\hbar\omega = 1.54$\,eV and $\phi_{0}=0$. The phase-matching is adjusted for each energy. Blue circles: Initial wavepacket as in Fig.~\ref{fig1}(c). Red squares: Plane wave. (b) The same as (a), but as a function of electric field amplitude for 100\,eV electrons. (c) Dependence of the trap \REV{width} on the relative phase between electron and light field for the wavepacket. \REV{(d) Dependence on the group velocity dispersion of the wavepacket, $\phi_0 = 0$.}}
\label{fig3}
\end{figure}

At this moment, we can control and manipulate the spectral confinement of the laser-modulated free electrons, enabling their trapping in the energy domain. This is achievable because the preceding analysis clarifies both the electron and light field parameters in each regime, as we will now demonstrate in detail. From now on throughout our study we stay in the Raman-Nath regime ($E_\mathrm{f} \approx 1\,\mathrm{V\,nm}^{-1}$ and $\hbar\omega=1.54$\,eV). For our spectral trap, it is reasonable to define an effective trap width \REV{$\Delta E_\mathrm{trap}$} as the energy difference between the minimum and the maximum populated sidebands \REV{(see Fig.~\ref{fig1}(c))}. \REV{Large widths $\Delta E_\mathrm{trap}$ correspond to weak trapping, whereas narrow $\Delta E_\mathrm{trap}$ allow only a small number of sidebands to be populated. $\Delta E_\mathrm{trap}$ stays the same for long interaction times and only} depends on the field parameters and the initial electron energy distribution. In Fig.~\ref{fig3}(a), we show the dependence of \REV{$\Delta E_\mathrm{trap}$} on the electron energy for the wavepacket and the plane wave from Fig.~\ref{fig1}(c) and (d), respectively, with the phase-matching adjusted to each energy. Not surprisingly, an increase in energy leads to weaker influence of the quadratic potential and therefore weaker trapping. A similar increase in width is reached by increasing the optical field strength, which enhances the hopping (see Fig.~\ref{fig3}(b)). The definition of $\kappa$ reveals another parameter that influences the trap width, the \REV{phase $\phi_0$ of the optical field at $t = 0$}. This dependence can be explained with an analogy to the classical harmonic oscillator, for which the largest energy transfer between the driving force and mass occurs when the phase difference (time difference) between the two is $\frac{\pi}{2}$. \REV{In our definitions, this phase difference corresponds to $\phi_0 = \frac{3\pi}{2}$, where we indeed find the maximum $\Delta E_\mathrm{trap}$ (see Fig.~\ref{fig3}(c)).} \REV{Figure~\ref{fig3}(d) shows the dependence of \REV{$\Delta E_\mathrm{trap}$} on the group velocity dispersion (GVD) of the electron wavepacket for $\phi_0 = 0$. As we move away from zero GVD, the trap width is increasing as the initial wavepacket stretches in time and starts to resemble a continuous wavepacket (see Supplemental Material~\cite{NoteSuppl} for details). We note that spectral trapping can also occur for larger electron energies than those treated in this work. However, the trap width will be on the order of the electron energy, making the effect hard to observe (see Supplemental Material~\cite{NoteSuppl}).}

In Figure~\ref{fig1}(c), we notice the appearance of spectral Airy-like fringes forming at intermediate and long evolution times in the vicinity of the trap edges, i.e., at the turning points for a confined and oscillatory localized wavepacket. These fringes are due to an interference pattern which emerges from the different phases accumulated by the set of initially occupied sideband states. The difference in phase accumulation by the different plane waves is strongest in the vicinity of the trap edges. A few oscillation cycles are required to develop a pronounced fringe pattern. This may be demonstrated by initializing the wavepacket closer to the quadratic potential minimum by altering the phase difference, which results in longer interaction times for the fringes to form and vice versa.

In conclusion, our work explores the spectral dynamics of the resonant interaction of a low-energy electron with a strong light field. With the help of analytical models and numerical simulations, we find new spectral features appearing as the interaction evolves. Particularly, we report a confinement of the electron spectrum induced by the curvature of the electron dispersion which is not accessible for fast electrons. Tunable with the help of the optical field parameters, effective trap widths ranging from few tens to few hundred electron volts can be achieved. Our findings can be highly useful for \REV{quantum coherent} electron manipulation. First, a confinement to only a few energy sidebands in the Bragg regime offers a natural truncation of the infinite Hilbert space of the multi-level Rabi ladder, allowing for adiabatic eliminations that enable us to get rid of undesired sidebands and higher-order photon transitions. Second, the finite Hilbert space as a synthetic spatial dimension can be then of aid for carrying out quantum simulations and computations utilizing coherent laser-electron interactions, as proposed recently for fast electrons~\cite{pan2022synthetic}. Finally, the dynamical confinement in the low-energy regime can help \REV{to control} the maximum energy transfer for laser-driven charged-particle acceleration, the minimum bunching size of an electron beam and the maximum photon energy produced in free-electron radiation.

The authors acknowledge Martin Koz\'ak, Bin Zhang, Ido Kaminer \REV{and Peter Hommelhoff} for useful discussions. This project has received funding from the European Union's Horizon 2020 research and innovation program under grant agreement No 853393-ERC-ATTIDA. We also acknowledge the Helen Diller Quantum Center at the Technion for \REV{partial financial} support.


%

\end{document}



\title{Self-trapping of slow electrons in the energy domain \\ -- Supplemental Materials --}

\author{Maor Eldar}
\affiliation{Department of Physics, Technion---Israel Institute of Technology, Haifa 32000, Israel}
\affiliation{Solid State Institute, Technion---Israel Institute of Technology, Haifa 32000, Israel}
\affiliation{\REV{The Helen Diller Quantum Center, Technion---Israel Institute of Technology, Haifa 32000, Israel}}

\author{\REV{Zhaopin Chen}}
\affiliation{Department of Physics, Technion---Israel Institute of Technology, Haifa 32000, Israel}
\affiliation{Solid State Institute, Technion---Israel Institute of Technology, Haifa 32000, Israel}
\affiliation{\REV{The Helen Diller Quantum Center, Technion---Israel Institute of Technology, Haifa 32000, Israel}}

\author{Yiming Pan}
\affiliation{Department of Physics, Technion---Israel Institute of Technology, Haifa 32000, Israel}
\affiliation{\REV{The Helen Diller Quantum Center, Technion---Israel Institute of Technology, Haifa 32000, Israel}}
\affiliation{School of Physical Science and Technology and Center for Transformative Science, ShanghaiTech University, Shanghai 200031, China}

\author{Michael Kr\"uger}
\altaffiliation{Corresponding author: krueger@technion.ac.il}
\affiliation{Department of Physics, Technion---Israel Institute of Technology, Haifa 32000, Israel}
\affiliation{Solid State Institute, Technion---Israel Institute of Technology, Haifa 32000, Israel}
\affiliation{\REV{The Helen Diller Quantum Center, Technion---Israel Institute of Technology, Haifa 32000, Israel}}

\date{\today}

\maketitle

\section*{Table of contents}
\begin{enumerate}[label=\Roman*.]
\item General tight-binding-approximated equation using the Floquet-Bloch theorem.
\item The Raman-Nath regime: multi-level Rabi model.
\item Slow electron energy-momentum conservation condition.
\item \REV{3D model of light-electron interaction and validity of 1D approaches.}
\item \REV{Proposed design of a double-grating DLA structure for modulating low-energy electrons.}
\item \REV{Full numerical solution of the 1D time-dependent Schr\"odinger equation.}
\item \REV{The role of the ponderomotive term in electron-light interactions.} 
\item Non-symmetric Bloch oscillations.
\item \REV{Spectral trapping for large electron or photon energies.}
\item \REV{Influence of the initial spectral phase of the electron wavepacket.}
\item \REV{Analogy to acousto-optic modulation.}
\end{enumerate}

\section{General tight-binding-approximated equation using the Floquet-Bloch theorem}

In accordance with our description of the slow-electron resonant interaction setup in the main text, we begin with the time-dependent Schr{\"o}dinger equation:

\begin{equation} \label{eq1}
i\hbar\frac{\partial}{\partial t}|\psi(t)\rangle=(\hat{H}_{0}+\hat{H}_\mathrm{I})|\psi(t)\rangle,
\end{equation} 
in which $|\psi(t)\rangle$ is the electron wavevector, $\hat{H}_{0}$ is the free-electron Hamiltonian and $\hat{H}_\mathrm{I}$ is the electron-light interaction Hamiltonian. A Taylor expansion of the relativistic energy dispersion to second order (see section 3 for derivation) leads to

\begin{equation} \label{eq2}
H_{0}=E_{0}+v_{0}(p-p_{0})+\frac{(p-p_{0})^{2}}{2\gamma^3 m}.
\end{equation} 
Notice that henceforth we omit operator signs for convenience. The electron's initial kinetic energy is $E_{0}=\sqrt{m^2c^4+p_{0}^2c^2} - mc^2$, where $p_0$ is its initial momentum, $c$ the vacuum speed of light, $\gamma$ is the Lorentz factor, $e$ is the elementary charge and $m$ is the electron's rest mass. The interaction Hamiltonian results from the minimal coupling by substituting $p$ by $p-eA$,
\begin{equation} \label{eq3}
H_\mathrm{I}=-\frac{e}{2\gamma m}\big[A\cdot p+ p\cdot A\big]+\frac{e^{2}A^{2}}{2\gamma m},
\end{equation} 
where for our concern the laser-induced vector potential along electron propagation direction is given by $A=A(z,t)=-\frac{E_\mathrm{f}}{\omega}\sin(\omega t-k_{z}z+\phi_{0})$. Here $E_\mathrm{f}$, $\omega$ and $\phi_{0}$ are the electric field amplitude, angular frequency and initial phase, respectively. $k_z$ is the optical wavevector along the propagation direction of the electron.

Considering the spatiotemporal periodicity of the vector potential, we employ the Floquet-Bloch ansatz:
\begin{align} 
|\psi(z,t)\rangle=\sum_{n=-\infty}^{\infty}a_{n}(t)e^{-i\omega_{n}t}|k_{n}\rangle ,\mathrm{\ or} \nonumber \\ \label{eq4} \psi(z,t)=\sum_{n=-\infty}^{\infty}a_{n}(t)e^{-i(w_{n}t-k_{n}z)},
\end{align} 
where we define
\begin{align*}
\omega_{n}=\omega_{0}+n\omega,\\ k_{n}=k_{0}+n k_{z}, \\ \omega_{0}=E_{0}/\hbar, \\ k_{0}=p_{0} \hbar.
\end{align*} 

We substitute Eq.~\ref{eq4} into Eq.~\ref{eq1}. First we derive the left-hand side:

\[
i\hbar\sum_{n}e^{-i(\omega_{n}t-k_{n}z)}\left(-i\omega_{n}+\frac{\partial}{\partial t}\right)a_{n}(t)=\sum_{n}e^{-i(\omega_{n}t-k_{n}z)}\left(\hbar \omega_{n}+i\hbar\frac{\partial}{\partial t}\right)a_{n}(t),
\]
and the right-hand side takes the form:
\[
\sum_{n=-\infty}^{\infty}\left(E_{0}+v_{0}(p-p_{0})+\frac{(p-p_{0})^{2}}{2\gamma^3 m} -\frac{e}{2\gamma m}\big[A\cdot p+p\cdot A\big]+\frac{e^{2}A^{2}}{2\gamma m}\right) a_{n}(t)e^{-i(\omega_{n}t-k_{n}z)}.
\]
We now evaluate each term of the right hand side individually. First, we calculate the $H_{0}$ term:
\begin{equation} \label{eq5}
H_{0}|\psi(t)\rangle=\sum_{n} \left( E_{0}+v_{0}n\hbar k_{z}+\frac{(n\hbar k_{z})^{2}}{2\gamma^3 m} \right) a_{n}(t) e^{-i(\omega_{n}t-k_{n}z)},
\end{equation}
where we used the de-Broglie relation, $p=\hbar k$. Next, we calculate the linear term $A\cdot p + p\cdot A$:
\[
H_\mathrm{I}|\psi(t)\rangle=-\frac{e}{2\gamma m}\left(A\cdot p + p\cdot A\right)\sum_{n}a_{n}(t)e^{-i\omega_{n}t}|k_{n}\rangle.
\]
Recalling $p=-i\hbar\frac{\partial}{\partial z}$ and $I=\int dz|z\rangle\langle z|$, we continue

\begin{align} 
H_\mathrm{I} |\psi(t)\rangle =&\frac{e}{2 \gamma m} (-A i\hbar\frac{\partial}{\partial z}-i\hbar\frac{\partial}{\partial z} A(z,t))\sum_{n=-\infty}^{\infty}a_{n}(t)e^{-i\omega_{n}t}\int dz|z\rangle\langle z|k_{n}\rangle \nonumber \\
=&-\hbar\frac{e A_{0}}{\gamma m} \sum_{n}[(k_{n}-i\frac{\partial}{\partial z})a_{n}(t)]e^{-i\omega_{n}t}\int dz\,e^{-i k_{n}z}\sin(\omega t-k_{z}z+\phi_{0})|z\rangle \nonumber \\
&-i\hbar\frac{e A_{0}}{2 \gamma m}k_{z}\sum_{n}a_{n}(t)e^{-i\omega_{n}t}\int dz\,e^{i k_{n}z}\cos(\omega t-k_{z}z+\phi_{0}), \label{eq6}
\end{align} 

where we labeled $A_{0}=-\frac{E_{f}}{\omega}$ and applied the projection $\langle z|k_{n}\rangle=e^{i k_{n}z}$. We now use the known decomposition:

\begin{align*}
\sin(\omega t-k_{z}z+\phi_{0})=-\frac{i}{2}[e^{i(\omega t-k_{z}z+\phi_{0})}-e^{-i(\omega t-k_{z}z+\phi_{0})}],\\
\cos(\omega t-k_{z}z+\phi_{0})=\frac{1}{2}[e^{i(\omega t-k_{z}z+\phi_{0})}+e^{-i(\omega t-k_{z}z+\phi_{0})}].
\end{align*}

By inserting this decomposition into Eq.~\ref{eq6} we obtain:

\begin{align*} 
H_\mathrm{I} |\psi(t)\rangle &=i\hbar\frac{e A_{0}}{\gamma m} \sum_{n}[(k_{n}-\frac{k_{z}}{2}-i\frac{\partial}{\partial z})a_{n}(t)]\int dz \, e^{-i(\omega_{n-1}t-k_{n-1}z-\phi_{0})}|z\rangle\\
&-i\hbar\frac{e A_{0}}{\gamma m} \sum_{n}[(k_{n}+\frac{k_{z}}{2}-i\frac{\partial}{\partial z})a_{n}(t)]\int dz \, e^{-i(\omega_{n+1}t-k_{n+1}z-\phi_{0})}|z\rangle.
\end{align*} 

Inserting the identity operator $I=\frac{1}{2\pi}\int d k|k\rangle\langle k|$ we find

\begin{align*} 
H_\mathrm{I} |\psi(t)\rangle  & = i\hbar\frac{e A_{0}}{\gamma m} \sum_{n}[(k_{n}-\frac{k_{z}}{2}-i\frac{\partial}{\partial z})a_{n}(t)]\frac{1}{2\pi} \int dz \int d k\, e^{-i k z} e^{-i(\omega_{n-1}t-k_{n-1}z-\phi_{0})}|k\rangle\\
&\ \ -i\hbar\frac{e A_{0}}{\gamma m} \sum_{n}[(k_{n}+\frac{k_{z}}{2}-i\frac{\partial}{\partial z})a_{n}(t)]\frac{1}{2\pi} \int dz \int d k\, e^{-i k z}e^{-i(\omega_{n+1}t-k_{n+1}z-\phi_{0})}|k\rangle\\
&= i\hbar\frac{e A_{0}}{\gamma m} \sum_{n}[(k_{n}-\frac{k_{z}}{2}-i\frac{\partial}{\partial z})a_{n}(t)]\frac{1}{2\pi} \int d k \left[ \int d z\, e^{-i(k-k_{n-1})z} \right] e^{-i(\omega_{n-1}t-\phi_{0})}|k\rangle\\
&\ \ -i\hbar\frac{e A_{0}}{\gamma m} \sum_{n}[(k_{n}+\frac{k_{z}}{2}-i\frac{\partial}{\partial z})a_{n}(t)]\frac{1}{2\pi} d k \left[ \int d z\, e^{-i(k-k_{n+1})z} \right] e^{-i(\omega_{n+1}t-\phi_{0})}|k\rangle\\
&=i\hbar\frac{e A_{0}}{\gamma m} \sum_{n}[(k_{n}-\frac{k_{z}}{2}-i\frac{\partial}{\partial z})a_{n}(t)]\frac{1}{2\pi} \int d k\, \delta(k-k_{n-1}) e^{-i(\omega_{n-1}t-\phi_{0})}|k\rangle\\
&\ \ -i\hbar\frac{e A_{0}}{\gamma m} \sum_{n}[(k_{n}+\frac{k_{z}}{2}-i\frac{\partial}{\partial z})a_{n}(t)]\frac{1}{2\pi} d k \int d k\, \delta(k-k_{n+1})e^{-i(\omega_{n+1}t-\phi_{0})}|k\rangle\\
&=i\hbar\frac{e A_{0}}{\gamma m} \sum_{n}[(k_{n}-\frac{k_{z}}{2}-i\frac{\partial}{\partial z})a_{n}(t)]\frac{1}{2\pi} e^{-i(\omega_{n-1}t-\phi_{0})}|k_{n\REV{-}1}\rangle\\
&\ \ -i\hbar\frac{e A_{0}}{\gamma m} \sum_{n}[(k_{n}+\frac{k_{z}}{2}-i\frac{\partial}{\partial z})a_{n}(t)]\frac{1}{2\pi} e^{-i(\omega_{n+1}t-\phi_{0})}|k_{n+1}\rangle.
\end{align*} 

Next, we shift the dummy index $n$ to find:

\begin{align*} 
H_\mathrm{I} |\psi(t)\rangle &=i\hbar\frac{e A_{0}}{\gamma m}\sum_{n}e^{-i\omega_{n}t}[a_{n+1}(t)(k_{n+1}-\frac{k_{z}}{2})e^{i\phi_{0}}-a_{n-1}(t)(k_{n-1}+\frac{k_{z}}{2})e^{-i\phi_{0}}]|k_{n}\rangle \\
&\ \ +\hbar\frac{e A_{0}}{\gamma m}\sum_{n}e^{-i\omega_{n}t}[e^{i\phi_{0}}\frac{\partial}{\partial z}a_{n+1}(z,t)-e^{-i\phi_{0}}\frac{\partial}{\partial z}a_{n-1}(z,t)]|k_{n}\rangle.
\end{align*} 

Assuming $a_{n}$ is slowly varying, we neglect second order derivatives and are thus left with:

\begin{equation} \label{eq7}
H_\mathrm{I} |\psi(t)\rangle \approx i\hbar\frac{e A_{0}}{\gamma m}\sum_{n}e^{-i\omega_{n}t}[a_{n+1}(t)(k_{n+1}-\frac{k_{z}}{2})e^{i\phi_{0}}-a_{n-1}(t)(k_{n-1}+\frac{k_{z}}{2})e^{-i\phi_{0}}]|k_{n}\rangle.
\end{equation}

Next, we derive the ponderomotive term $\frac{(e A)^{2}}{2 \gamma m}$:

\begin{align*} 
&\frac{e^{2}E_\mathrm{f}^{2}}{2\gamma m \omega^{2}}\sum_{n}(\sin(\omega t-k_{z}z+\phi_{0}))^{2}a_{n}(t)e^{-i(\omega_{n}t-k_{n}z)}\\
&=\frac{e^{2}E_\mathrm{f}^{2}}{2\gamma m \omega^{2}}\sum_{n}\left\{\frac{e^{2i(\omega t-k_{z}z+\phi_{0})}+e^{-2i(\omega t-k_{z}z+\phi_{0})}-2}{-4}\right\}e^{-i(\omega_{n}t-k_{n}z)}a_{n}(t)\\
&=-\sum_{n} \left[ \frac{e^{2}E_\mathrm{f}^{2}}{8\gamma m \omega^{2}}\{e^{-i(\omega_{n-2}t-k_{n-2}z-2\phi_{0})}+e^{-i(\omega_{n+2}t-k_{n+2}z+2\phi_{0})}\}-\frac{e^{2}E_\mathrm{f}^{2}}{4\gamma m \omega^{2}}e^{-i(\omega_{n}t-k_{n}z)} \right] a_{n}(t).
\end{align*} 

In the first two terms we shift the dummy index $n$: $n\rightarrow n\pm2$ to finally find:
\begin{equation} \label{eq8}
=-\frac{e^{2}E_{f}^{2}}{4\gamma m \omega^{2}}\sum_{n}[\frac{1}{2}\{a_{n+2}(t)e^{i2\phi_{0}}+a_{n-2}(t)e^{-i2\phi_{0}}\}-a_{n}(t)]e^{-i(\omega_{n}t-k_{n}z)}.
\end{equation}

Substituting the results given by Eq.~\ref{eq5}, Eq.~\ref{eq7} and Eq.~\ref{eq8}, we obtain the general tight-binding-approximated equation of motion:

\begin{align}
\begin{split}\label{eq9}
&\left[i\hbar\frac{\partial}{\partial t}+n\hbar(\omega-v_{0}k_{z})-n^{2}\frac{(\hbar k_{z})^{2}}{2m}+U_\mathrm{p}\right]a_{n}(t) \\
&=\delta\left((k_{n+1}-\frac{k_{z}}{2})a_{n+1}(t)e^{i\phi}+(k_{n-1}+\frac{k_{z}}{2})a_{n-1}(t)e^{-i\phi}\right) \\
&\ \ -\frac{U_\mathrm{p}}{2}\left(a_{n+2}(t)e^{i2\phi_{0}}+a_{n-2}(t)e^{-i2\phi_{0}}\right),
\end{split}
\end{align}

where $\phi=\phi_{0}+\frac{\pi}{2}$. We stress that the linear on-site potential term $n\hbar(\omega-v_{0}k_{z})$ stems from the phase mismatching, and the quadratic on-site potential term $n^{2}\frac{(\hbar k_{z})^{2}}{2m}$ originates from the electron dispersion curvature. The nearest-neighbor coupling term $\delta=e\hbar E_\mathrm{f}/(2m\omega)$ is laser induced, and $\pm \frac{k_{z}}{2}$ originates from the gauge contribution of $\nabla\cdot A\neq 0$. Lastly, the next-nearest-neighbor coupling showing the two-photon scattering process is proportional to the ponderomotive energy $U_\mathrm{p}=e^{2}E_\mathrm{f}^{2}/{(4m\omega^{2})}$.\\

We recall that in Wannier space which describes PINEM-type spectra we found the simplified equation of motion:

\begin{align} 
i\hbar\frac{\partial}{\partial t}a_{n}=(\kappa a_{n+1}+\kappa^{*}a_{n-1})+\beta n^{2}a_{n}. \label{eq10}
\end{align}

 with the parameters $\kappa, \beta$.  From this tight-binding Schr{\"o}dinger equation, we wish to find the Hamiltonian, therefore we first span the slowly rotating wavevector using the PINEM ladder basis:
\[
|\psi(t)\rangle=\sum_{n}a_{n}(t)|n\rangle.
\]
\REV{Thus,} we can simply read off our tight-binding Hamiltonian directly from Eq.~\ref{eq10}:

\begin{equation}\label{eq11}
\hat{H}=\sum_{n}\beta n^{2}|n\rangle\langle n|+(\kappa|n\rangle\langle n+1|+c.c).
\end{equation}

\section{The Raman-Nath regime: multi-level Rabi model}

We recall the form of Eq.~\ref{eq10} as the general tight-binding-approximated equation: 
\[
i\hbar\dot{a}_{n}(t)=\epsilon n^{2}a_{n}+\kappa a_{n+1}-\kappa^{*}a_{n-1},
\]
where $\epsilon=\frac{p_{z}^{2}}{2m}$ is the energy of a plane wave with momentum $p_{z}$. In the Raman-Nath regime we find $|\kappa|\gg\epsilon$. Therefore when treating $\epsilon$ as a perturbation, the unperturbed equation reads:

\begin{equation}\label{eq18}
i\hbar\dot{a}_{n}(t)=\kappa a_{n+1}-\kappa^{*}a_{n-1}.
\end{equation}

We now wish to solve for Eq.~\ref{eq18}, to do so we apply \REV{a} phase transformation on Eq.~\ref{eq18}: $\tilde{a}_{n}(t)=a_{n}e^{in(\phi_{0}+\frac{\pi}{2})}$ \REV{and rewrite $\kappa=|\kappa|e^{i(\phi_{0}+\frac{\pi}{2})}$ in accordance with Eq.~\ref{eq9}}:
\[
|\kappa|e^{i(\phi_{0}+\frac{\pi}{2})}a_{n+1}=|\kappa|e^{i(\phi_{0}+\frac{\pi}{2})}\tilde{a}_{n+1}e^{-i(\phi_{0}+\frac{\pi}{2})}e^{-in(\phi_{0}+\frac{\pi}{2})}=|\kappa|\tilde{a}_{n+1}e^{-in(\phi_{0}+\frac{\pi}{2})},
\]
\[
|\kappa|e^{-i(\phi_{0}+\frac{\pi}{2})}a_{n-1}=|\kappa|e^{-i(\phi_{0}+\frac{\pi}{2})}\tilde{a}_{n-1}e^{i(\phi_{0}+\frac{\pi}{2})}e^{\REV{-}in(\phi_{0}+\frac{\pi}{2})}=|\kappa|\tilde{a}_{n-1}e^{\REV{-}in(\phi_{0}+\frac{\pi}{2})}.
\]
Thus by \REV{invoking the phase transformation and} multiplying Eq.~\ref{eq18} by $e^{in(\phi_{0}+\frac{\pi}{2})}$ we find
\[
i\hbar\dot{\tilde{a}}_{n}=|\kappa|(\tilde{a}_{n+1}\REV{-}\tilde{a}_{n-1}).
\]
\begin{equation}\label{eq19}
-\frac{i\hbar}{|\kappa|}\dot{\tilde{a}}_{n}=(\tilde{a}_{n-1}-\tilde{a}_{n+1}).
\end{equation}
From Eq.~\ref{eq19} we find the probability amplitude coefficients uphold the Bessel function recurrence relation
\begin{equation*}
J_{n-1}-J_{n+1}=2J_{n}',
\end{equation*}
i.e.,
\begin{equation*}
\tilde{a}_{n}=J_{n} \left( \frac{2|\kappa|}{i\hbar}t \right).
\end{equation*}

Finally, we obtain the expression for a typical PINEM-type spectrum,
\begin{equation}\label{eq20}
\boldsymbol{a_{n}=J_{n} \left( \frac{2|\kappa|}{i\hbar}t \right) e^{in\phi_{0}}}.
\end{equation}

\section{Slow electron energy-momentum conservation condition}

We begin by Taylor expanding the relativistic electron dispersion to second order:
\begin{equation}\label{eq21}
E=\sqrt{c^{2}p^{2}+(m c^{2})^{2}}.
\end{equation}

Here $E,p,m$ are electron total energy, momentum and mass, respectively, and $c$ is the speed of light. We first evaluate first and second derivatives of $E$ at the initial momentum (velocity) $p_{0}\ (v_{0})$: 
\[
\frac{\partial E}{\partial p}|_{p_{0}}=\frac{2c^{2}p}{2\sqrt{c^{2}p^{2}+(mc^{2})^{2}}}|_{p_{0}}=\frac{c^{2}p_{0}}{E_{0}}=\frac{c^{2}\gamma m v_{0}}{E_{0}}=v_{0}.
\]
\[
\frac{\partial^{2}E}{\partial p^{2}}=c^{2}(\frac{1}{E}-p\frac{1}{E^{2}}\frac{\partial E}{\partial P})=c^{2}(\frac{1}{E}-p\frac{1}{E^{2}}\frac{c^{2}p}{E})=c^{2}(\frac{1}{E}-p\frac{v}{E^{2}}).
\]
\[
\rightarrow\frac{\partial^{2}E}{\partial p^{2}}|_{p_{0}}=c^{2}(\frac{1}{\gamma mc^{2}}-\gamma mv_{0}\frac{v_{0}}{(\gamma m c)^{2}})=\frac{1}{\gamma m}(1-(\frac{v}{c})^{2})=\frac{1}{\gamma^{3}m}.
\]

We obtain the Taylor expansion
\begin{equation*}
E\sim E_{0}+v_{0}(p-p_{0})+\frac{(p-p_{0})^{2}}{2m\gamma^{3}}+\mathcal{O}((p-p_{0})^{3})\ \ \mathrm{for}\ \ p\rightarrow p_{0},
\end{equation*}
where $\gamma$ is the Lorentz factor, $\gamma=\frac{1}{\sqrt{1-\beta}}$, and $\beta=\frac{v}{c}$.

Recalling from our Floquet-Bloch ansatz that $k_{n}=k_{0}+n k_{z}$, we can rewrite $p-p_{0}$ as:
\begin{equation}\label{eq22}
E=E_{0}+v n\hbar k_{z}+\frac{(v n\hbar k_{z})^{2}}{2\gamma^{3}m}.
\end{equation}

\textbf{From Eq.~\ref{eq22} we clearly see the physical origin of the quadratic potential. It emerges from the finite curvature of the electronic dispersion which is accessible for slow phase-matched electrons.}

Park et al.~\cite{park2010photon} found that the energy-momentum conservation constraint for electron-electric field interactions manifests as

\[
\cos(\theta_\mathrm{c})=\frac{2\hbar\omega E+(\hbar\omega)^{2}-(\hbar kc)^{2}}{2\hbar kc^{2}p},
\]

where $\hbar k$ is the recoil momentum magnitude and $\theta_\mathrm{c}$ is the critical angle, defined such that energy conservation at the single photon exchange between the light field and the electron is maintained, i.e. $E+\hbar\omega=\sqrt{c^{2}(\vec{p}+ \Vec{p}_\mathrm{photon})^{2}+(mc^{2})^{2}}$ holds. Here the photon momentum is $\vec{p}_\mathrm{photon}=(\hbar k \cos(\theta),\hbar k \sin(\theta))$. For a swift electron, e.g. $E=200$\,keV, the interaction with a field with photon energy $\hbar\omega=1.54$\,eV yields $\frac{E}{\hbar\omega}\approx10^{5}$. Thus, $2\hbar\omega E\gg(\hbar\omega)^{2}-(\hbar kc)^{2}$ and we
find the known energy-momentum conservation condition for the fast electron:

\begin{equation}\label{eq23}
\frac{2\hbar\omega E}{2\hbar kc^{2}p}=\frac{v_{p}}{v} \leq 1,
\end{equation}

where $v_{p}$ is the optical field's phase velocity. Thus, conservation of energy and momentum has been reduced to matching the photon phase velocity and electron group velocity.

For a slow electron, e.g. $E=100$\,eV, interacting with the same field we find $\frac{E}{\hbar\omega}\approx10^{2}$. Therefore we keep the additional terms to find a correction for the slow electron energy-momentum conservation condition,
\[
\cos(\theta_\mathrm{c})=\frac{2\hbar\omega E+(\hbar\omega)^{2}-(\hbar kc)^{2}}{2\hbar kc^{2}p}=\frac{2\hbar\omega E}{2\hbar kc^{2}p}+\frac{(\hbar\omega)^{2}}{2\hbar kc^{2}p}-\frac{(\hbar kc)^{2}}{2\hbar kc^{2}p}.
\]
Invoking the field phase velocity and electron group velocity we find

\begin{equation}\label{eq24}
{\cos(\theta_\mathrm{c})=\frac{v_{p}}{v}\left(1+\frac{\hbar\omega}{2E}\left\{1-\left(\frac{c}{v_{p}}\right)^{2}\right\}\right)}.
\end{equation}

Equation~\ref{eq24} reveals that for the correction to be significant the electron's initial kinetic energy must be comparable to the photon energy due to the prefactor $\frac{\hbar\omega}{2E}$. Thus even for our slow electron with initial kinetic energy of 100\,eV the correction remains negligible. 

\section{3D model of light-electron interaction and validity of 1D approaches}

\REV{In the following, we formulate an approach for solving the problem of the interaction of a slow electron with light in 3D and discuss its relation to the 1D theory approaches described in the main text. We will discuss how the 1D theory follows from the 3D approach and the validity of this simplification.}

\REV{We start by considering the relativistic energy dispersion $E_{0}=\sqrt{m^2c^4+p_{0}^2c^2}$ of the electron wavepacket and expand it to the second-order as $E = E_0 + \textbf{v}_0 (\textbf{p} - p_0 \textbf{e}_z) + \frac{p_x^2}{2 \gamma m} + \frac{p_y^2}{2 \gamma m} + \frac{(p_z - p_0)^2}{2 \gamma^3 m}$, where $E_0$ denotes the center energy of the initial wavepacket, $p_0$ is the initial momentum along the $z$ direction and $\textbf{v}_0$ is the group velocity. Notice that the Lorentz factor is $\gamma \approx 1$ for the slow-electron coupling in our setup. Correspondingly, we can write the time-dependent Schr\"odinger equation of a slow electron in the presence of an electromagnetic field as follows:}

\REV{\begin{equation} \label{eq3D1}
    i\hbar \left( \frac{\partial}{\partial t} + \textbf{v}_0 \cdot \nabla \right) \psi = 
    -\frac{(\partial_x^2 + \partial_y^2 + \partial_z^2)}{2m} \psi - (e \textbf{A} \cdot \textbf{v}_0 )\psi + \frac{e^2 A^2}{2m} \psi,
\end{equation}}

\REV{where the wavefunction $\psi(x,y,z,t)$ is the slowly varying part of the full electron wavefunction $\Psi=\psi(x,y,z,t) \exp[-(i E_0 t)/\hbar + (ip_0 z)/\hbar]$. Here, we apply the minimal coupling with the substitution $\textbf{p} \rightarrow \textbf{p}-e\textbf{A}$ to involve the laser field through its vector potential $\textbf{A}$. The last term in Eq.~\ref{eq3D1} is the ponderomotive modulation, which is quadratic in the field strength of the laser. Considering the electron propagating along the $z$ axis, $\textbf{v}_0=(0,0,v_0 )$, we obtain}

\REV{\begin{equation} \label{eq3D2}
    i\hbar \left( \frac{\partial}{\partial t} + v_0 \frac{\partial}{\partial z} \right) \psi = 
    -\frac{(\partial_x^2 + \partial_y^2)}{2m} \psi - \frac{1}{2m}\partial_z^2 \psi - \left( e v_0 A_z(x,y,z,t) 
    + \frac{e^2 A_z^2(x,y,z,t)}{2m}\right) \psi.
\end{equation}}

\REV{Here the left-hand side (LHS) of the equation denotes the center-of-motion dynamics, which corresponds to a co-moving frame $\zeta = z-v_0 t$. The right-hand side (RHS) of the equation is decomposed into three terms. The first term represents the dispersion dynamics in the transverse directions $x$ and $y$, while the second term represents the dispersion in the longitudinal direction, providing the harmonic potential to trap the energy-modulated electron in the energy domain. The third term corresponds to the linear and the ponderomotive laser modulations, depending on the electron's propagation direction. For the resonant light-electron interaction, the linear modulation dominates over the ponderomotive modulation, allowing us to neglect the latter in the main manuscript.}

\REV{This is an appropriate 3D TDSE model of our slow-electron resonant interaction with light. \REVFINAL{We note that a full numerical 2D or 3D solution of our problem coupled to a Maxwell field solver requires significant computing resources and time on the week to month scale, which goes beyond the scope of this work.} Notice that for our concern in the main text, the first term (transverse dynamics) is disregarded in our 1D model. However, the 1D model neglects the transverse dynamics and the inhomogeneity of the laser field, which may be capable of affecting the physics in our 3D model. In order to quantify these effects, we express the inhomogeneous vector potential strength as}

\REV{\begin{equation}
    A_z (x,y,z,t) = A_0 (x,y) \cos(\omega_\mathrm{L} t - k_z z + \phi_0 ),
\end{equation}}

\REV{in which the amplitude of the vector potential $A_0 (x,y)$ carries the inhomogeneous transverse profile. In DLA structures, such inhomogenous fields cannot be avoided. Suppose that the slow-electron wavefunction is given by $\psi=f_t (x,y)\varphi(z,t)$, where $f_t (x,y)$ denotes the transverse profile and $\varphi(z,t)$ the longitudinal profile. This separation of variables is valid when the transversal beam size is smaller than the field variation of the inhomogeneous field. If this condition is not met, a full solution of the 3D TDSE is required.}

\REV{For the DLA realization of our scheme, one may employ a single grating illuminated by an impinging electric field with wavelength $\lambda=800$\,nm, a slow electron with 100\,eV kinetic energy and a grating period of $\lambda_z = 16$\,nm. These parameters lead to an evanescent mode inside the DLA which decays exponentially in the direction transverse to the electron propagation direction with $1/e$ decay length $\Gamma \sim \lambda_z / (2\pi) \sim 2.5$\,nm. In comparison, the de-Broglie wavelength of our slow electron is $\sim$0.12\,nm and a typical beam size in electron microscopy is $\sim$1\,nm. This suggests that the transverse effects persist in this configuration. However, there are two ways to overcome these effects and maintain validity and predictions of our 1D model. One possibility is the inclusion of an identical second grating to the DLA realization of our scheme, resulting in a mirror-symmetric double grating structure, where an electron propagates in a channel between two gratings facing each other. This allows for a nearly homogeneous transverse field distribution in the center of the channel between the two gratings (cf.~\cite{Peralta2013}). Another possibility is to exploit the near field's exponential decay profile in the transverse dimension by simply initializing the electron's trajectory at a larger distance from the grating (cf.~\cite{Li2023}). This would also result in a nearly homogeneous transverse field distribution. The two approaches have been demonstrated experimentally~\cite{Peralta2013,Li2023} and show that the experimental realization of our work is feasible. In the next section, we will detail a possible double-grating structure.}

\section{Proposed design of a double-grating DLA structure for modulating low-energy electrons}

\REV{In this section, we describe a double-grating DLA structure which is suitable for realizing a spatial harmonic optical mode synchronized with the motion of slow electrons. We choose a double grating structure (cf.~Ref.~\cite{Peralta2013}) with simultaneous illumination from both sides in order to ensure a nearly homogeneous field in the transverse direction. The grating period required for 100\,eV electrons is $\lambda_z = 15.8$\,nm. The other grating dimensions and the material are less important and only influence how efficiently the incident laser field amplitude is translated into the strength of the first spatial harmonic mode created by the gratings. Here we choose fused silica as the grating material and the channel width between the teeth of the grating as 15\,nm, with the electron traversing the center of the channel. The height of each tooth is 22.5\,nm. We calculated the electromagnetic fields in the structure using a Maxwell solver (Ansys Lumerical), assuming an illumination of 800\,nm light from both sides and a structure with infinite periodicity.}

\begin{figure} [h]
\begin{centering}
\includegraphics[width=0.65\columnwidth]{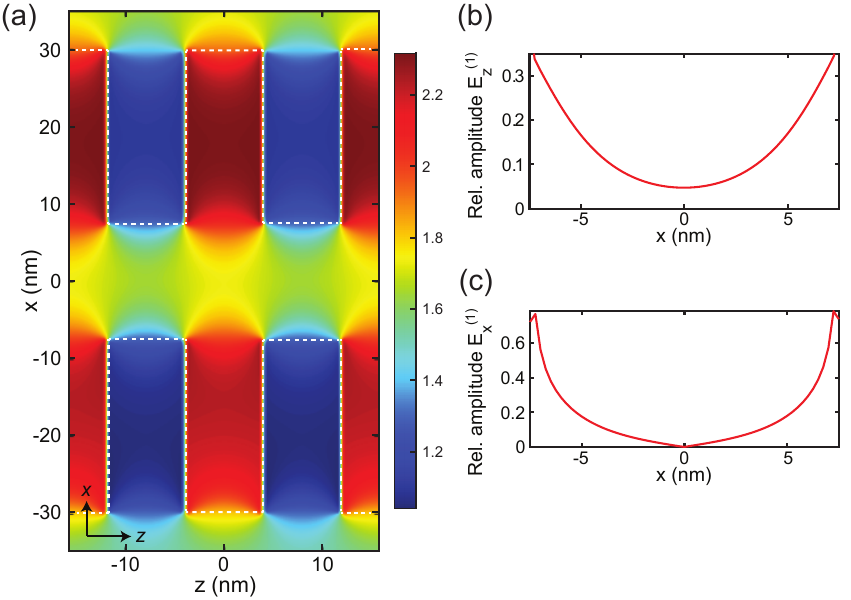}
\par\end{centering}
\caption{\REV{Proposed design for a suitable DLA structure. (a) Temporal snapshot of the longitudinal field component $E_z$ in units of the incident field strength. Inside the channel, a spatial modulation of $E_z$ with the grating periodicity is clearly visible, which leads to the optical mode driving the modulation of incident electron pulses. The dashed white lines indicate the grating boundaries. (b) Transverse profile of the longitudinal field amplitude of the first spatial harmonic mode, $E_z^{(1)}$, relative to the incident field. Within 1\,nm from the center, the variation of the amplitude is less than 5\%, allowing for a nearly homogeneous modulation of a narrow electron beam. (c)  Transverse profile of the transverse field component of the first spatial harmonic mode, $E_x^{(1)}$, relative to the incident field.}}
\label{fig:Sgrating}
\end{figure}

\REV{Figure~\ref{fig:Sgrating}(a) shows a temporal snapshot of the longitudinal field component $E_z$ at the peak of the laser field. Inside the channel, a spatial modulation of $E_z$ with the grating periodicity is clearly visible, which leads to the optical mode driving the modulation of incident electron pulses. The longitudinal field component of this mode (first spatial harmonic) along $z$ can be described as $E_z^{(1)}(x,z) = 2 \delta_z(x) E_\mathrm{f}^\mathrm{(in)} \cos{(\omega t - k_z z + \phi_0})$, where $E_\mathrm{f}^\mathrm{(in)}$ is the incident field amplitude, $k_z = 2\pi / \lambda_z$ is the grating wavevector and $\delta_z(x)$ is an efficiency factor. Now we can identify the optical field amplitude $E_\mathrm{f}$ from the main text with $2 \delta_z(x) E_\mathrm{f}^\mathrm{(in)}$. The factor of two accounts for the fact that the light is incident from two sides (double illumination). We obtain $\delta_z(x = 0)$ in the center of the structure as 5\%, which means that 5\% of the incident light amplitude is leading to phase-matched electron modulation. This factor is quite low due to the evanescent nature of the spatial harmonic mode. Higher-order modes are strongly suppressed in this geometry. Generating an optical mode with amplitude $0.5\,\mathrm{V\,nm}^{-1}$ (cf.~Fig.~1(c) and (d) in the main text) is experimentally feasible and is compatible with the damage threshold of fused silica of $10\,\mathrm{V\,nm}^{-1}$ for suitable laser parameters~\cite{Soong2012}. Moreover, the transverse profile of $\delta_z(x)$ (see Fig.~\ref{fig:Sgrating}(b)) shows that the field is quite homogeneous within a distance of 1\,nm from the center axis, ensuring a nearly homogeneous interaction with a narrow electron beam.}

\REV{We note that our 3D model described in the previous section does not account for deflection, i.e., a change of direction of the electron trajectories away from the longitudinal direction. We anticipate some deflection with slow electrons and strong fields due to the breakdown of the non-recoil approximation and due to the presence of transverse electric field components in the DLA scheme. The latter can be seen in Fig.~\ref{fig:Sgrating}(c) where we plot the amplitude of the {\em transverse} field component $E_x^\mathrm{(1)}$ of the first spatial harmonic as a function of $x$. In the very center, there are no transverse forces on the electron. However, deflecting fields are increasing as we move away from the center. Within 1\,nm distance from the center, $E_x^\mathrm{(1)}$ remains significantly lower than the $E_z^\mathrm{(1)}$, hence deflection effects can be mitigated for a sufficiently narrow electron beam. Our work represents an initial exploration, and further investigations are needed to understand the transverse beam direction in the context of low-energy electrons and strong fields.}

\REVFINAL{Our approach does not only require the size of the electron beam to be narrow, but also its divergence to be sufficiently small. For the case of the Raman-Nath regime calculation of Fig.~1(c) and (d), for instance, we require a beam divergence angle smaller than 6\,mrad to remain in an effective 1D system. The electron beam of laser-triggered nanotip sources has a much larger divergence so that electron optics is required to refocus the beam to a small spot. Because the source size of laser-triggered nanotips can be smaller than 1\,nm~\cite{Ehberger2015}, such focusing is feasible.}

\section{Full numerical solution of the 1D time-dependent Schr\"odinger equation}

\REV{In order to further verify our results, we carried out a full numerical integration of the time-dependent Schr\"odinger equation (TDSE) and compared it with the full Floquet-Bloch model (Eq.~1 of the main paper. We employed a 4th order Runge-Kutta algorithm to integrate the TDSE (time step 6.7\,as, spatial grid step 2.4\,\AA), with $E_\mathrm{f}=0.5\,\mathrm{V\,nm}^{-1}$, $\hbar \omega=1.54\,$eV and $\phi_{0}=0$. Instead of a plane wave (cf.~Fig.~1(c)), we assumed a realistic finite energy width of 0.1\,eV, corresponding to an electron wavepacket duration much longer than the optical cycle duration of 800\,nm light. The result is shown in Fig.~S2(a) and (b). All dynamical spectral features of the Floquet-Bloch model result (Fig.~S2(c) and (d)) are also found in the TDSE result. Furthermore, the small asymmetry between energy gain and loss mentioned in the main text can be observed in (b) and (d) at $t = 12.5$\,fs. }

\begin{figure} [h]
\begin{centering}
\includegraphics[width=0.65\columnwidth]{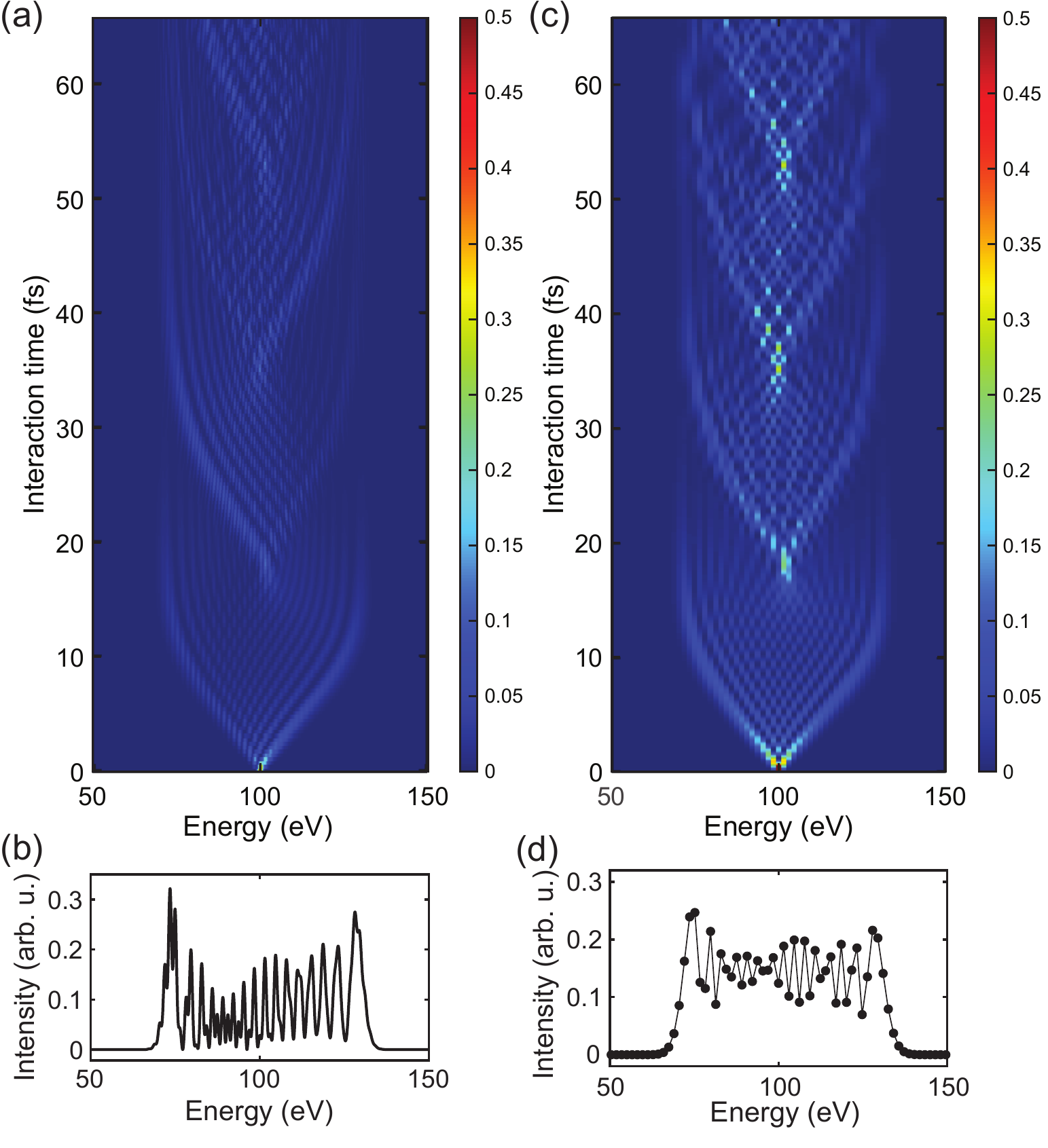}
\par\end{centering}
\caption{\REV{Comparison between the solution of the time-dependent Schr\"odinger equation (TDSE) and the Floquet-Bloch model (Eq.~1 in the main text). (a) TDSE calculation for an initial wavepacket with a spectral width of 0.1\,eV, much smaller than the photon energy of the driving light field. (b) Spectrum for $t = 12.5$\,fs. (c) Spectral evolution calculated with the Floquet-Bloch model. (d) Spectrum for $t = 12.5$\,fs. We note that the Floquet-Bloch model gives data points only at the discrete sideband energies.}}
\end{figure}

\section{The role of the ponderomotive term in electron-light interactions}

\REV{In the first part of this section} we theoretically study the ratio between the two terms for different field strengths interacting with a slow electron. To this end, we employ the simpler non-relativistic dispersion. The interaction between the electron and the electric field is given by the following Hamiltonian:

\begin{equation}\label{eq25}
H=\frac{p^{2}}{2m}+\frac{e p A}{m}+\frac{e^{2}A^{2}}{2m}.
\end{equation}

We compare the $p \cdot A$ and ponderomotive term contributions for a slow electron with 50\,eV kinetic energy and an extremely slowly moving electron with 5\,eV energy. We examine the following field strengths: $0.09-10^{3}\,\mathrm{V\,nm}^{-1}$. Our choice is motivated by commonly used and accessible field strength in typical experiments: $0.09-2\,\mathrm{V\,nm}^{-1}$ are used in swift electron PINEM experiments, $10^{2}\,\mathrm{V\,nm}^{-1}$ is the required field strength for ionizing hydrogen, and $10^{3}\,\mathrm{V\,nm}^{-1}$ is employed in plasma physics. In addition, from Eq.~\ref{eq25} we find: $\frac{e^{2}A^{2}}{2m_{e}}/\frac{epA}{m_{e}}=\frac{eA}{p}$.  Therefore, in order to distinguish the dominant term we focus on the quantity $\frac{eA}{p}$. 

\begin{center}
\begin{tabular}{ |p{3cm}|p{3cm}|p{3cm}|  }
 \hline
  \multicolumn{3}{|c|}{${e A}/{p}$} \\
 \hline
$E_\mathrm{f}\,(\,\mathrm{V\,nm}^{-1})$ & $E=50\,\mathrm{eV}$ & $E=5\,\mathrm{eV}$\\
 \hline
 $0.09$   & $1.6*10^{-3}$ & $5*10^{-3}$ \\
 \hline
 $2$      & $3.5*10^{-2}$ & $1.1*10^{-1}$\\
  \hline
 $10$     & $1.7*10^{-1}$ & $1.12$\\
  \hline
 $10^{2}$ & $1.7$         & $1.12*10$\\
  \hline
 $10^{3}$ & $1.77*10$     & $1.12*10^{2}$\\
 \hline
\end{tabular}
\end{center}

We find that for a 50\,eV electron, even at field strengths on the order of $10^{3}\,\mathrm{V\,nm}^{-1}$ the ponderomotive term is only larger by an order of magnitude compared to the $p \cdot A$ term. At the atomic ionization field strength $10^{2}\,\mathrm{V\,nm}^{-1}$ they are of equal contribution, and for $1\,\mathrm{V\,nm}^{-1}$ the ponderomotive term can be treated as a perturbation. In addition, we find that as the electron becomes slower, e.g. at 5\,eV kinetic energy which is on the order of the photon energy of 1.54\,eV, lower field strengths are needed for the ponderomotive term to have stronger contribution. For example, for a 5\,eV electron a relatively weak field strength of $0.1\,\mathrm{V\,m}^{-1}$ already elevates the ponderomotive term to the extent of a perturbation which is two orders of magnitude weaker then the field strength required for the same effect for a 50\,eV electron.

\REV{In the second part of this discussion, we will briefly compare our scheme, which relies on the $A\cdot p$ interaction, to two other approaches. The first approach is a ponderomotive (two-photon) interaction, where phase mismatch causes highly asymmetric spectra~\cite{kozak2022asynchronous}. The second study introduces an inelastic Kapitza-Dirac interaction involving a bi-chromatic laser field and an auxiliary harmonic potential, which is general and can be applied not only to electrons, but also atoms and molecules~\cite{Huang2021}. Both have in common that they are based on ponderomotive two-photon interactions, which usually necessitates stronger fields (at least by one order of magnitude) and longer interaction times (picoseconds instead of tens of femtoseconds) compared to our scheme because of their use of $A^2$ term. Furthermore, while both the phase-mismatched interaction and the auxiliary harmonic potential can lead to a confinement and an effective truncation of the Hilbert space of the interaction, a strong confinement down to an effective two-level system can only be reached with our phase-matched interaction in the Bragg regime with the help of the parabolic trapping potential inherently associated with the slow-electron dispersion.}

\section{Non-symmetric Bloch oscillations}

From Eq.~\ref{eq9} we find that when there is phase \REV{mismatch}, i.e. $\hbar\omega \neq \hbar v_{0}k_{z}$, the dominant on-site potential is the linear one. Therefore, the electron experiences a constant ``field'' in the energy domain leading to the emergence of Bloch-like oscillations. We find the oscillations are inherently non-symmetric for the slow electron from a simulation (see Fig.~\REV{S3}).

\begin{figure} [h]
\begin{centering}
\includegraphics[width=0.65\columnwidth]{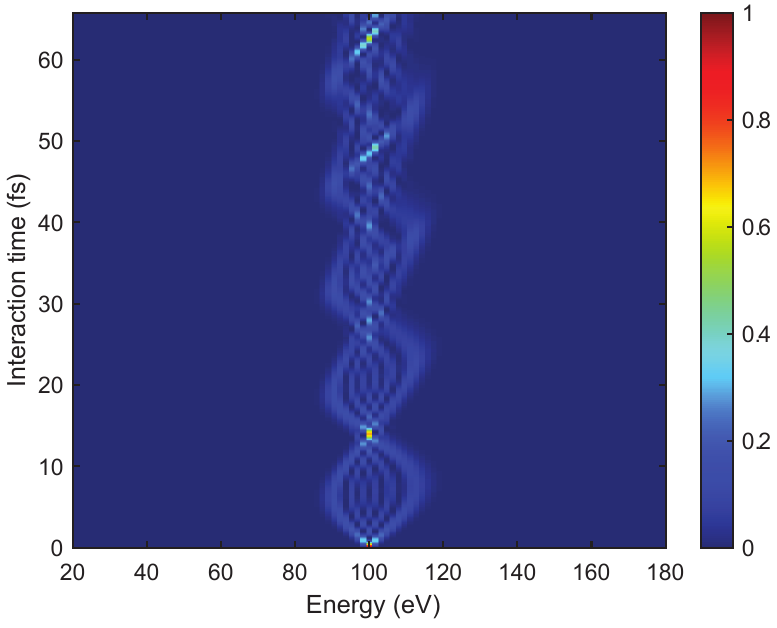}
\par\end{centering}
\caption{Slow electron non-symmetric Bloch oscillations in the energy domain. Field parameters: \REV{$E_\mathrm{f}=0.5\,\mathrm{V\,nm^{-1}}$}, $\phi_{0}=0$, $\hbar\omega=1.54$\,eV. The electron's kinetic energy is 100\,eV. A DLA grating with a period length of $\lambda_\mathrm{p}\approx 23$\,nm introduces phase mismatching.}
\end{figure}

From the simulation we clearly see an asymmetry evolving with time between absorption and emission side bands. This asymmetry emerges as the maximal occupied absorption band is larger then the corresponding emission band. This effect leads to a certain time delay between the time period of a single oscillation for absorption and emission bands.

\REV{We find that energy confinement is stronger for the mismatched case ($\sim$\,40\,eV) than for the phase-matched case ($\sim$\,80\,eV). However, these electron-field parameters correspond to the Raman-Nath regime, where confinement ($\propto \beta_\mathrm{d}$) is somewhat suppressed compared to the nearest neighbor hopping ($\propto \kappa$). Therefore, it is only with the phase-matched trapping that we can reach the Bragg regime which allows for ultra-strong confinement of only two energy levels. The phase-matched case is preferred for trapping over the phase-mismatched case. We use the phase-matched energy trapping to create a finite Hilbert space, so we can manipulate the superposition of the sidebands and further shape the wavefunction of the slow electron. With the truncation of the infinite Hilbert space, it holds promise for quantum simulation or quantum information processing with the slow light-modulated electrons.}

\section{Spectral trapping for large electron or photon energies}

\REV{Our simulations include calculations with a maximum initial kinetic energy electron of 1\,keV interacting with 1.54\,eV photons. Here the trap width is considerable, $\Delta E \approx E_0/2$, meaning the energy confinement becomes comparable to the electron's initial energy. This trend increases with increasing electron initial kinetic energies, as seen in Fig.~3(b) in the main text. In most previous studies, only fast electrons ($v > 0.1c$) were employed. There, the kinetic energies ranged from 10\,keV (e.g., in a scanning electron microscopy setup) to 200\,keV (transmission electron microscope setup) combined with infrared photons. However, the confinement effect was not observed in these cases, as the energy gain/loss induced in these light-electron interactions did not reach the trap width predicted in our work. For instance, in~\cite{Dahan2021} the gain was 100\,eV for 200\,keV electrons. The underlying reason for this lack of confinement is that the relativistic dispersion for fast electrons is practically linear, resulting in a very small curvature. In contrast, our study reveals that the amplitude of the confining quadratic potential well in the electron energy domain ($\beta_\mathrm{d}$) is, in fact, the curvature of the electronic relativistic dispersion. This disparity is the fundamental cause of the practical absence of confinement for fast electrons in previous experiments. }

\REV{This explanation stands also for the case of the photon energy on the order of the electron energy. If we employ sub-relativistic (100\,keV) electrons interacting with extreme X-ray photons ($\sim$\,100\,keV), the phase matching condition (equal field phase velocity and electron velocity) will be strongly violated after the electron has absorbed or emitted just a single photon. Consequently, the quadratic potential becomes negligible compared to the linear potential, whose amplitude is proportional to this phase mismatch, leading to effective Bloch oscillation dynamics. Such dynamics were predicted previously in~\cite{pick2018bloch} and~\cite{pan2022synthetic} for fast electrons interacting with ultraviolet photons in a non-resonant interaction. }

\section{Influence of the initial spectral phase of the electron wavepacket}

\REV{In the main text, we explore the dependence of the trap width on the electron wavepacket’s spectral phase distribution, which is indeed a crucial factor. Here we consider a quadratic spectral phase, which is nothing else than the group velocity dispersion (GVD). High absolute values of the GVD stretch the wavepacket temporally. As expected, increasing the GVD in absolute values increases the electron wavepacket’s trap width towards the value for the single-energy state (see Fig.~3(d) in the main text). We define the spectral phase as a function of GVD $D_2$ in the following way:}

\REV{\begin{equation}
    \phi_n = \frac{D_2}{2} n^2 \omega^2,
\end{equation}}

\REV{where $n$ is the number of the Floquet-Bloch sideband.}

\REV{We also consider the effect of a random phase, with $E_\mathrm{f}=0.5\,\mathrm{V\,nm}^{-1}$, $\hbar\omega = 1.54$\,eV, $\phi_{0}=0$, $\Delta E\sim 7.5$\,eV. We vary the phase of the sidebands randomly and sum over the evolution plots resulting from Eq.~1 in the main text incoherently. The result is shown in Fig.~S4. We find that the sum over random phases recovers the plane wave case shown in Fig.~1(d) in the main text.}

\begin{figure} [h]
\begin{centering}
\includegraphics[width=0.65\columnwidth]{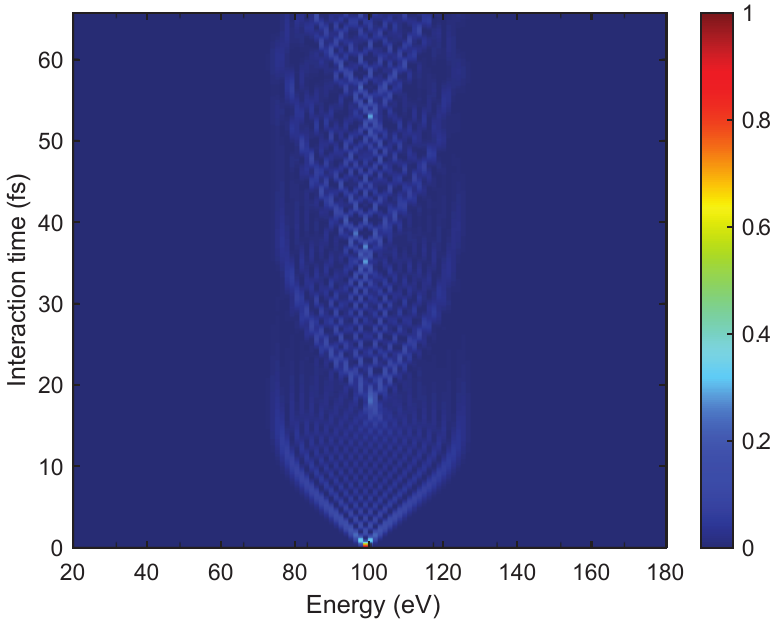}
\par\end{centering}
\label{fig:Srandom}
\caption{\REV{Wavepacket evolution with random phases. See the text for more details.}}
\end{figure}

\section{Analogy to acousto-optic modulation}

\REV{The simplified Eq.~2 in the main text shows that the resonant slow-electron-light interaction can be understood within the Raman-Nath formalism that was originally devised for acousto-optics~\cite{Korpel1981}. In this section, we will show this analogy more explicitly. The Raman-Nath parameter is known in acousto-optics as quality or Klein-Cook factor $Q$ given by}

\REV{\begin{equation}
\label{eq:aomQ}
Q = 2\pi\frac{\lambda L}{n \Lambda^2},
\end{equation}}

\REV{where $\lambda$ is the wavelength of incoming light diffracted by an acoustic wave with wavelength $\Lambda$ inside a medium of longitudinal interaction length $L$~\cite{Korpel1981}. For $Q \ll 1$, acousto-optic interaction in the Raman-Nath regime leads to multiple diffraction of the incoming light, described by a Bessel function. In contrast, the Bragg regime $Q \gg 1$ leads to efficient Bragg diffraction into a single diffraction order, which is the mechanism used in the widely used acousto-optic modulator (AOM). Equation~\ref{eq:aomQ} shows that the length of the medium $L$ governs the acousto-optic interaction regime. A thick medium enables Bragg diffraction, whereas a thin medium favors Raman-Nath diffraction.}

\REV{A closer look to the definition of the Raman-Nath parameter $\rho$ for the slow-electron-light interaction reveals the analogy to acousto-optics. Assuming $\phi = 3\pi/2$ without the loss of generality, we can express $\rho$ differently as}

\REV{
\begin{equation}
\rho = \frac{\hbar^2 k_z^2 \omega}{\hbar k_0 e E_\mathrm{f}} = 2 \pi \frac{\lambda_\mathrm{dB} L_\mathrm{eff}}{\lambda_z^2},
\end{equation}
}

\REV{where $\lambda_\mathrm{dB}$ is the de-Broglie wavelength of the electron which experiences modulation by the light field mode created by a grating with period $\lambda_z$, in full analogy to $\lambda$ and $\Lambda$ from acousto-optics. We also find an effective interaction length $L_\mathrm{eff}$ which is given by}

\REV{
\begin{equation}
\label{eq:Leff}
L_\mathrm{eff} = \frac{\hbar \omega}{e E_\mathrm{f}}.
\end{equation}
}

\REV{Here we may interpret $L_\mathrm{eff}$ as the distance over which a single light quantum $\hbar \omega$ is absorbed or emitted by the electron. We can find that this interpretation makes sense by taking a look at the Raman-Nath regime results in Fig.~1(d) in the main text performed with $\hbar\omega = 1.54$\,eV and $E_\mathrm{f} = 0.5\,\mathrm{V\,nm}^{-1}$. At a kinetic energy of 100\,eV, the electron has a velocity of $5.94\,\mathrm{nm\,fs}^{-1}$. Until time $t \sim 7.5$\,fs, the electron gains 20\,eV or 13 photon energies. We can then estimate the distance for gaining one photon energy as 3.5\,nm, which is very close the result of Eq.~\ref{eq:Leff}, $L_\mathrm{eff} = 3.1$\,nm.}

\REV{While Bragg-like diffraction is both achieved in our scheme with electrons and in acousto-optics, the reflections at the trap edges observed in our case cannot be readily observed in acousto-optics. The underlying reason is that the acousto-optical modulation manipulates the light beam using transverse acoustic waves, resulting in transverse momentum transfer, while in our case the electron beam is modulated longitudinally, resulting in longitudinal momentum transfer. We can therefore reach a regime where the continuous interaction pushes the electron wave out of the phase-matching condition and the interaction has the opposite effect, throwing the electron back towards its original state. In the extreme case, this leads to a two-level system with Rabi-like oscillations.}


%